\documentclass{aa520}
\usepackage{graphicx}

\begin{document}
  \title{Photometric redshifts with the $Multilayer$ $Perceptron$ Neural
Network: application to the HDF--S and SDSS}
\titlerunning{Photometric redshifts with a MLP Neural Network}
   \author{Vanzella E.\inst{1,2}
        \and
        Cristiani S.\inst{3}
        \and
        Fontana A.\inst{4}
        \and
        Nonino M. \inst{3}
        \and
    Arnouts S.\inst{5}
        \and
    Giallongo E.\inst{4}
        \and
        Grazian A.\inst{4}
        \and
        Fasano G.\inst{6}
    \and
    Popesso P.\inst{7}
    \and
    Saracco P.\inst{8}
    \and
    Zaggia S.\inst{3}
        }
   \offprints{E.Vanzella (vanzella@pd.astro.it)}
   \institute{European Southern Observatory, Karl-Schwarzschild-Str. 2,
   D-85748 Garching, Germany
   \and
        Dipartimento di Astronomia dell'Universit\`a di Padova,
        Vicolo dell'Osservatorio 2,
        I-35122 Padova, Italy
   \and
        INAF - Osservatorio Astronomico di Trieste, Via G.B. Tiepolo 11,
        40131 Trieste, Italy
   \and
        INAF - Osservatorio Astronomico di Roma, via
        dell'Osservatorio 2, Monteporzio, Italy
   \and
    Laboratoire d'Astrophysique de Marseille,
    Traverse du Siphon-Les trois Lucs, 13012 Marseille, France
   \and
    INAF - Osservatorio Astronomico di Padova, Vicolo
        Osservatorio 5 Padova, Italy
   \and
    Max-Planck-Institut fur extraterrestrische Physik,
    D-85740, Garching, Germany
   \and
    INAF - Osservatorio Astronomico di Brera, via Brera 28
    20121, Milano, Italy}
   \date{Received \dots; accepted \dots}

   \abstract{
We present a technique for the estimation of photometric redshifts
based on feed-forward neural networks. The {\it Multilayer
Perceptron} (MLP) Artificial Neural Network is used to predict
photometric redshifts in the HDF--S from an ultra
deep multicolor catalog.
Various possible approaches for the training of the neural network are
explored, including the deepest and most complete spectroscopic redshift catalog
currently available (the Hubble Deep Field North dataset) and models
of the spectral energy distribution of galaxies available in the
literature. The MLP can be trained on observed data, theoretical data and
mixed samples.
The prediction of the method is tested on the spectroscopic sample in
the HDF--S (44 galaxies). Over the entire redshift range, $0.1<z<3.5$,
the agreement between the photometric and spectroscopic redshifts in the
HDF--S is good: the training on mixed data produces
$\sigma_{z}^{test} \simeq 0.11$,
showing that model libraries together with observed data
provide a sufficiently complete description of the galaxy population.
The neural system capability is also tested in a low redshift regime, $0<z<0.4$,
using the Sloan Digital Sky Survey
Data Release One (DR1) spectroscopic sample.
The resulting accuracy on 88108 galaxies is
$\sigma_{z}^{test} \simeq 0.022$.
Inputs other than
galaxy colors - such as morphology, angular size and surface
brightness - may be easily incorporated in the neural network
technique. An important feature, in view of the application of the
technique to large databases, is the
computational speed: in the evaluation phase, redshifts of $10^5$
galaxies are estimated in few seconds.

}
   \maketitle
%
%________________________________________________________________

\section{Introduction}

Deep multicolor surveys, using a selection of broad- and/or
intermediate-band filters to simultaneously cover the spectral energy
distribution (SED) of a large number of targets, have been an
important part of astronomy for many years but have remarkably surged
in popularity in recent times. Digital detectors and telescopes with
improved spatial resolution in all wavelength regimes have enabled
astronomers to reach limits that were unthinkable only a few decades
ago and are now revealing extremely faint sources (see for a review
Cristiani, Renzini \& Williams 2001).
A general hindrance for the transformation of this wealth of data into
cosmologically useful information is the difficulty in obtaining
spectroscopic redshifts of faint objects, which, even with the new
generation of 8m-class telescopes, is typically limited to
I(AB)$\simeq$25.  This has spurred a widespread interest in the
estimation of the redshift directly from the photometry of the targets
(photometric redshifts). Major spectral features, such as the $Balmer$
$Break$ or the $Lyman$ $limit$, can be identified in the observed SED
and, together with the overall spectral shape, make possible a redshift
estimation and a spectral classification.

The photometric redshift techniques described in the literature can be
classified into two broad categories: the so-called empirical training
set method, and the fitting of the observed Spectral Energy
Distributions by synthetic or empirical template spectra. In the first
approach (see, for example, \cite{connolly95}), an empirical
relation between magnitudes and redshifts is derived using a subsample
of objects in which both the redshifts and photometry are available
(the so-called {\it training set}).
A slightly modified version of this method was used by \cite{wang98}
to derive redshifts in the HDF--N by means of a linear
function of colors.

In the SED-fitting approach a spectral library is used to compute
the colors of various types of sources at any plausible redshift,
and a matching technique is applied to obtain the ``best-fitting''
redshift.  With different implementations,
this method has been used in the HDF--N (\cite{rv02}, \cite{mass01}, 
\cite{sawicki97}, \cite{soto99}, \cite{benitez00}, \cite{arno99}) and ground--based
data (\cite{giallo00}, \cite{font99}, \cite{font00}).
%., Pello et al. 1999).

A crucial test in all cases is the comparison between the
photometric and spectroscopic redshifts which is typically limited
to a subsample of relatively bright objects.

In the present work, photometric redshifts have been obtained using
a $Multilayer$ $Perceptron$ Neural Network (MLP) with the primary
goal of recovering the correct redshift distributions up to the
highest redshifts in deep fields such as the HDFs.
The method has been tested on the HDF--S spectroscopic
sample (0.1$<$z$<$3.5) and on a sample of galaxies
(in a relatively low-redshift regime 0$<$z$<$0.4)
from the Sloan Digital Sky Survey
Data Release One (SDSS DR1, \cite{abaza03}).

The structure of this paper is as follows: in Section 2 we give an
introduction to the neural network methods. Section 3 describes the
training set for the HDF--S and Section 4 the training technique. In
Section 5 we apply the method to the spectroscopic sample in the
HDF--S. An application to the SDSS DR1 samples is
described in section 6.
Section 7 is dedicated to a general discussion. Our conclusions are
summarized in Section 8.

\section {Artificial Neural Networks}

According to the $DARPA~Neural~Network~Study$ (1988, AFCEA
International Press),
a neural network is a system composed of many simple processing elements
operating in parallel whose function is determined by the network structure,
connection strengths, and the processing
performed at the computing elements or nodes.

An artificial neural network has a natural proclivity for
storing experimental knowledge and making it available for use. The
knowledge is acquired by the network through a learning process and
the interneuron connection strengths - known as synaptic weights - are
used to store the knowledge (\cite{haykin94}).

There are numerous types of neural networks (NNs) for addressing
many different types of problems, such as modelling memory,
performing pattern recognition, and predicting the evolution of
dynamical systems. Most networks therefore perform some kind of
data modelling.

The two main kinds of learning algorithms are: $supervised$ and
$unsupervised$.  In the former the correct results (target values)
are known and given to the NN during the training
so that the NN can adjust its weights to try to match its outputs to the
target values. In the latter, the NN is not provided with the correct
results during training. Unsupervised NNs usually perform some kind of
data compression, such as dimensionality reduction or clustering.

The two main kinds of network topology are {\it feed-forward} and
{\it feed-back}. In feed-forward NN, the connections between units do not
form cycles and usually produce a relatively quick response to an input. Most
feed-forward NNs can be trained using a wide variety of efficient
conventional numerical methods (e.g. conjugate gradients,
Levenberg-Marquardt, etc.) in addition to algorithms invented by NN
researchers. In a feed-back or recurrent NN, there are cycles in the
connections.
In some feed-back NNs, each time an input is presented,
the NN must iterate for a potentially long time before producing a
response.

\subsection{The Multilayer Perceptron}

In the present work we have used one of the most important types
of supervised neural networks, the {\it
feed-forward~multilayer~perceptron} (MLP), in order to produce
photometric redshifts.  The term $perceptron$ is historical, and
refers to the function performed by the nodes. An introduction on
Neural Networks is provided by \cite{sarle94a}, and on multilayer
Perceptron by \cite{bailer01} and \cite{sarle94b}. A comprehensive
treatment of feed-forward neural networks is provided by
\cite{bishop95}.

\begin{figure}
 \includegraphics[width=90mm]{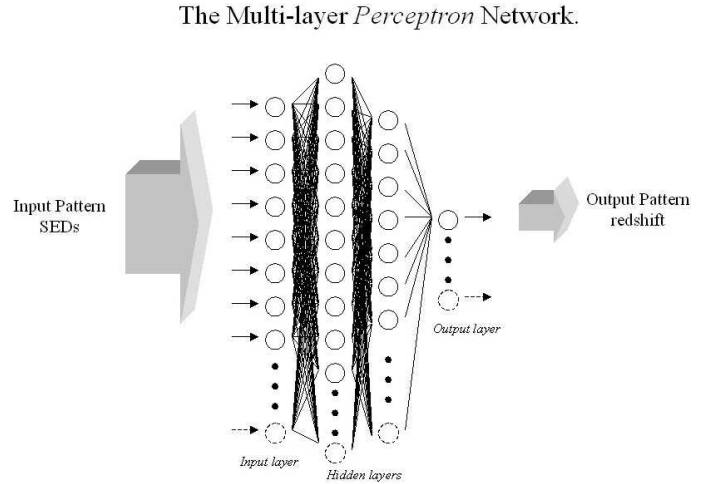}
 \caption{A general scheme of a multilayer Perceptron feed-forward neural
network.
\label{fig:MLP_scheme}}
  \end{figure}

In Fig.~\ref{fig:MLP_scheme} the general architecture of a network
is shown.  The network is made up of layers and each layer is
fully connected to the following layer. The layers between the
input and the output are called hidden layers and the
correspondent units, hidden units.

For each input pattern, the network produces an output pattern through
the $propagation$ $rule$, compares the actual output with the
desired one and computes an error. The learning algorithm adjusts the
weights of the connections by an appropriate quantity to reduce the
error ({\it sliding down the slope}). This process continues until the error
produced by the network is low, according to a given criterion (see
below).

\subsubsection{The propagation rule}
An input of a node ($net_{j}$) is the combination of the output of the
previous nodes ($o_{i}$) and the weights of the corresponding links ($w_{ij}$),
%in our case
the combination is linear: $net_{j} = \sum_{i} w_{ij}o_{i}$. Each
unit has a transform function (or activation function), which
provides the output of the node as a function of the $net$.
Nonlinear activation functions are needed to introduce
nonlinearity into the network.
We have used the {\it logistic} (or sigmoid) function: $out = 1
/ [1 + exp(-K net)]$ and the $tanh$ function $out$=$tanh(K net)$, for all
units. $K$ is the gain parameter fixed
before the learning. By increasing $K$ the activation function approximates a
step.
The propagation rule, from the input layer to the output layer, is
a combination of activation functions.

No significant difference has been found in the training
process between using the {\it logistic} and $tanh$ functions.

\subsubsection{Back-propagation of the error}
The weights, $\bf{w}$, are the free parameters of the
network and the goal is to minimize the total error function
with respect to $\bf{w}$ (maintaining a good generalization power, see below).

The error function in the weight space defines the multi-dimensional
error surface and the objective is to find the global (or acceptable
local) minima on this surface. The solution implemented in the present work
is the $gradient~descent$, within which the weights are adjusted
(from small initial random values) in order to follow the
steepest downhill slope. The error surface is not known in advance,
so it is necessary to explore it in a suitable way.

The error function typically used is the sum-of-squares error,
which for a single input vector, $n$, is

\begin{equation}
e^{\{n\}} = \frac{1}{2}\sum_i \beta_i(y_i^{\{n\}} - T_i^{\{n\}})^2
\label{eqno1}
\end{equation}

\noindent where $y_i$ is the output of the NN and $T_i$ is the target output
value for
the $i^{th}$ output node and $n$ runs form 1 to the total number of examples 
in the training set. In the present work $i$=1, a single output node is used to
estimate the redshift (other nodes could be used to estimate other quantities,
such as the spectral type).
The $\beta_i$ terms make it possible to assign different weights to
different outputs, and thereby give priority to the correct
determination of certain outputs.  In the gradient descent process the
weight vector is adjusted in the negative direction of the
gradient vector,
\begin{equation}
  \Delta {\bf w} = - \eta \frac{\partial e}{\partial {\bf w}}
\label{eqno2}
\end{equation}
and the new generic weight is
\begin{displaymath}
w_{new}= w_{old}+\Delta w
\end{displaymath}

The amplitude of the step on the error surface is set by the
$\eta$-learning parameter: large values of $\eta$ mean large steps.
Typically $\eta$ belongs to the interval (0,1], in this application a small
value has been used ($<0.005$) together with a high value of the gain in the
activation functions ($K=5$). If $\eta$ is too small the training time
becomes very long, while a large value can produce oscillations around
a minimum or even
lead to miss the optimal minimum in the error surface.

The learning algorithm used in the present work is the standard
$back-propagation$. It refers to the method for computing the gradient
of the case-wise error function with respect to the weights for a
feed-forward network.  $``Standard~backprop''$ is a definition of the
$generalized$ $delta$ $rule$, the training algorithm that was
popularized by Rumelhart, Hinton, and Williams in chapter 8 of
Rumelhart and McClelland (1986), which remains one of the most widely
used supervised training methods for neural nets.

This learning algorithm implies that the error function is continuous
and derivable, so that it is possible to calculate the gradient. For
this reason the activation functions (and their final combination
through the propagation rule) must be continuous and derivable. From the
computational point of view, the derivative of the activation
functions adopted in the present work is easily related to the value
of the function $out=F(net)$ itself (see Sec. 2.1.1: $F' \propto out(1-out)$
in the case $F=sigmoid$ or $F' \propto (1-out^{2})$ if $F=tanh$.)

 \begin{figure}
 \includegraphics[width=90mm]{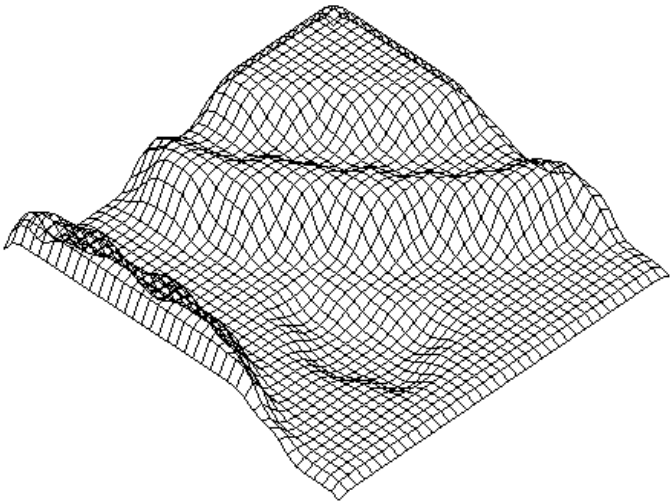}
 \caption{A simplified representation of the error surface: the behavior of the error
as a function of 2 weights. The momentum term improves the minimization during the training phase.
Momentum allows a network to respond to the local gradient and also to take into account
of the recent trends in the error surface. Acting like a low-pass filter, momentum allows
the network to ignore small features in the error surface. Without momentum a network
may get stuck in a shallow local minimum. With momentum a network can slide through
such a minimum.}
\label{fig:local_minim}
  \end{figure}

When the network weights approach a minimum solution, the gradient
becomes small and the step size diminishes too, giving origin to a
very slow convergence.
Adding a momentum (a residual of the previous weight variation)
to the equations of the weight update, the minimization improves
(\cite{bishop95}):

\begin{equation}
 w_{new} =  w_{old} + \Delta w + \alpha \Delta w_{old}
\label{eqno3}
\end{equation}
where $\alpha$ is the momentum factor (set to 0.9 in our
applications). This can reduce the decay in learning updates and
cause the learning to proceed through the weight space in a fairly
constant direction. Besides a faster convergence to the minimum,
this method makes it possible to escape from a local minimum if
there is enough momentum to travel through it and over the
following hill (see Fig.~\ref{fig:local_minim}). The generalized
delta rule including the $momentum$ is called the $``heavy~ball
~method''$ in the numerical analysis literature (\cite{berts}).

The learning algorithm has been used in the so called on-line (or
$incremental$) version, in which the weights of the connections are
updated after each example is processed by the network.  One $epoch$
corresponds to the processing of all examples one time.
The other possibility
is to compute the training in the so called $batch~learning$ (or epoch
learning), in which the weights are updated only at the end of each
epoch (not used in the present application).

\section{The training technique}
During the learning process, the output of a supervised neural net comes to
approximate the target values given the inputs in the training
set. This ability may be useful in itself, but more often the purpose
of using a neural net is to generalize, i.e. to
get some output from inputs
that are $not$ in the training set ($generalization$).
NNs, like other flexible nonlinear estimation methods
such as kernel regression and smoothing splines, can suffer from
either under fitting or over fitting. A network that is not sufficiently
complex\footnote[1]{The complexity of a network is related to both the
number of weights and the amplitude of the weights (the mapping produced by a
NN is an interpolation of the training data, a high order fit to data is
characterized by large curvature of the mapping function, which in turn
corresponds to large weights).}
can fail to fully detect
the signal in a complicated data set, leading to under fitting:
an {\it inflexible} model will have a large $bias$.
On the other hand a
network that is too complex may fit the noise, not just the signal,
leading to over-fitting: a model that is too flexible in
relation to the particular data set will produce a large $variance$,
(\cite{sarle95}).
The best generalization is obtained when the best compromise between these two
conflicting quantities (bias and variance) is reached.
There are several approaches to avoid under- and over-fitting,
and obtain a good generalization.
Part of them aim to {\it regularize}
the complexity of the network during the training phase,
such as the $Early~Stopping$ and $weight-decay$ methods
(the size of the weights are tuned in order to produce a mapping
function with small curvature, the large weights are penalized.
Reducing the size of the weights reduces also the ``effective'' number of
weights (\cite{moody92})).

A complementary technique belongs to the Bayesian framework, in which the
bias-variance trade off is not so relevant, and networks with high complexity
can be used without producing over-fitting (an example is to train a
$committee$ of networks, \cite{bishop95}).

\subsection{Generalize error}

\subsubsection{Early stopping}

The most commonly used method for estimating the generalization error
in neural networks is to reserve part of the data as a $test~set$,
which must not be used in $any$ way during the training. After the
training, the network is applied to the test set, and the error on the
test set provides an unbiased estimate of the generalization error,
provided that the test set was chosen in a random way.

In order to avoid (possible) over-fitting during the training, another
part of the data can be reserved as a $validation~set$ (independent
both of the training and test sets, not used for updating the weights),
and used during the training to monitor the generalization error.
The best epoch corresponds to the lowest validation error, and the
training is stopped when the validation error rate ``starts to go up''
($early~stopping$ method).
The disadvantage of this technique is that it reduces the amount of
data available for both training and validation, which is particularly
undesirable if the available data set is small.
Moreover, neither the training nor the
validation make use of the entire sample.

\subsubsection{Committees of networks}

As mentioned in the previous sections, an over-trained NN tends to
produce a large variance in the predictions maintaining a relatively
small bias. A method that reduces the variance (and keeps small the
bias) is to use a {\it committee} of NNs (\cite{bishop95}).
Each member of the committee differs from the other members for
the different training history. We have generated the members
using a bootstrap process, varying:

\begin{enumerate}
\item{the sequence of the input patterns (the
{\it incremental learning} method used in the present work
is dependent on the sequence presented).}

\item{the initial distribution of weights (the starting point
on the error surface).}

\item{the architecture of the NN (number of nodes and layers).}

\end{enumerate}

The final prediction, adopted
in the present work, is the mean and the median of the predictions
obtained from the members of the committee (with 1-$\sigma$ error or
16 and 84 percentiles).
Averaging over many solutions means reducing the variance.
Since the complexity of the individual member is not a problem,
the trainings have been performed without regularization and
at the lowest training-error the weights have been
frozen and used for the prediction.

This method has displayed a
better and stable generalization power with respect to a single training
(also using the validation set to regularize the learning).
Moreover this method gives a robust estimate of the error
bounds for the output of the network.

For these reasons the training described in the next sections has
been carried out using a committee of networks.

\section{The training-set}

Since we are using a supervised neural network, we need a
training-set.  Each element ({\it example}) in the training-set is
composed of a pair of vectors: the input pattern and the
target. For our purposes the input pattern contains the Spectral
Energy Distribution (SED) of the objects (but other configurations are
possible:
templates with a priori knowledge, SED plus the apparent luminosity in a
reference band, the angular size, the morphology, etc.).
The target in this application is the redshift.

The training
has been tested on the available spectroscopic sample in the HDF--S
({\cite{cristiani00}},
\cite{Rigopoulou00}, \cite{vanzella02}, Glazebrook et
al., {\tt http://www.aao.gov.au/hdfs/Redshifts/}).
The prediction of the redshifts in the HDF--S have been computed
following different approaches:

\begin{enumerate}
\item{training on the HDF--N spectroscopic sample using the colors
as an input pattern.} \item{training on the HDF--N spectroscopic
sample using the colors and the apparent luminosity in the I band
as an input pattern.} \item{training on both HDF--N spectroscopic
sample and a set of templates obtained from CWW (Coleman, Wu \&
Weedman) and/or from Rocca-Volmerange and Fioc (labelled RV00
hereafter).}
\item{Training on the CWW or RV00 SEDs alone (without
spectroscopic redshifts) have also been tested.}
\end{enumerate}

The photometry of the HDF--N has been obtained from the available catalog
provided
by \cite{soto99} whereas the photometric catalog of the HDF--S is
provided by \cite{vanzella01} and \cite{fontana03}.

\begin{figure}
 \includegraphics[width=90mm]{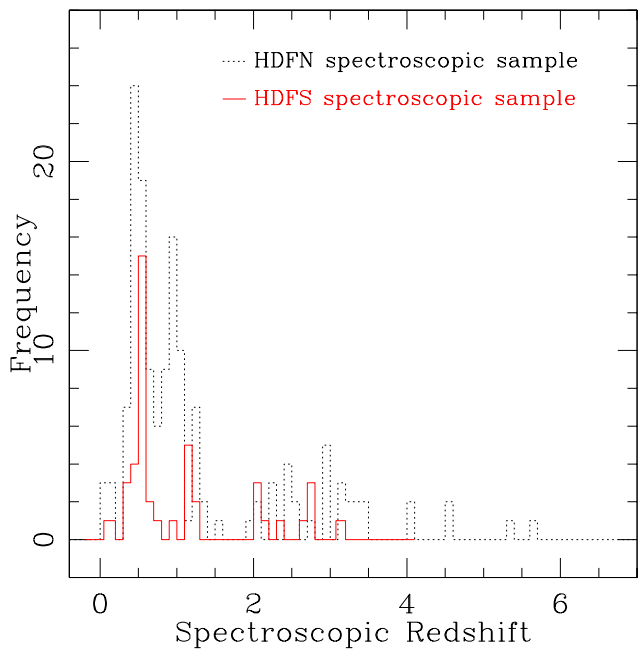}
 \caption{Spectroscopic redshift distributions of the two fields HDF--N (dashed
line) and HDF--S (solid line).
\label{fig:z_distib}}
  \end{figure}

The sample in the HDF--N contains 150 spectroscopic redshifts
(\cite{cohen00}, \cite{dawson01}, \cite{soto01}), while the sample
in the HDF--S contains 44 spectroscopic redshifts (in
Fig.~\ref{fig:z_distib} the redshift distributions of both fields
are shown).

In order to test the prediction we have used the variance as a statistical estimator:
$$ \sigma_{z}^{2} = \frac{1}{N}\sum_i (zNN_i - zspec_i)^{2} \ \ . \eqno(4) $$
where $zNN$ is the neural prediction, $N$ is the number of galaxies,
and $i$=1..$N$.
In the literature another statistical estimator is sometimes used, the
mean absolute deviation normalized by the $(1+z)$ factor
(e.g. \cite{labbe03}):
$$ \delta_{z} = \frac{1}{N}\sum_i \frac{|zNN_i - zspec_i|}{1 + zspec_i} \ \ .
\eqno(5) $$
The quantity $\delta_{z}$ has the advantage to be roughly uniform,
while the variance tends to increase with increasing redshift.

\subsection{The input pattern}

The magnitudes of the observed objects in a given photometric system are
the input of the network. In the present work the filters are $F300$, $F450$,
$F606$,
$F814$ (WFPC2, HST) and $Js$, $H$, $Ks$ for the near infrared (ISAAC,
VLT).
If the flux in a given band has a signal to noise ratio
less than 2.0 it is considered an upper limit in that band, and the value of
the flux is set to 1$\sigma$ error.

It is convenient to avoid too large input values that could cause a
saturation in the output of the activation functions (sigmoid or $tanh$),
but it is not necessary to rescale the inputs rigorously in the interval
[-1,1].
A non linear rescaling of the input 
values is also useful to make more uniform the function that the network 
is trying to approximate.

In the present application the input values have been rescaled:
$p_{i} = -0.5+[f_{i} / f_{F814}]^{0.4}$,
where $i$
runs over the following bands: $F300$, $F450$, $F606$, $Js$, $H$, $Ks$
and $f_{F814}$ is the flux in the reference $F814$ band.
When the apparent AB magnitude in the $F814$ band, $m_{814}$, is
used as an input (e.g. Sec. 5.1.2 and 5.2.2), it has been normalized 
as follow:

{\small
$$p_{F814} = \left[ \frac{1}{(m_{max}-m_{min})}
\right]([m_{814}-m_{min}]-[m_{max}-m_{814}])$$}
where $m_{max}$ is 28 and $m_{min}$ is 18.

\section{Redshift prediction on the HDF--S}

\small
\begin{table*}
%\onecolumn
\centering
\caption{Training of different architectures on the HDF--N
spectroscopic sample (150 objects) and evaluation on the
HDF--S spectroscopic sample. The number of epochs is 5000, the
bootstrap has been computed on 100 extractions (100 members of the
committee).}
\begin{tabular}{lccc|cccc}
\hline \hline
\multicolumn{4}{c|}{$Colors$ as } & \multicolumn{4}{c}{$Colors$ $\&$ $Magnitudes$}\cr
\multicolumn{4}{c|}{input pattern}& \multicolumn{4}{c}{as input pattern}\cr
\hline
 [Net]$_{\_Weights}$   &$<\sigma_{z}^{train}>$ &$\sigma_{z}^{test}$ &$\delta_{z}^{test}$  &[Net]$_{\_Weights}$&$<\sigma_{z}^{train}>$&$\sigma_{z}^{test}$&$\delta_{z}^{test}$\\
                       &                       & median/mean        &median/mean          &                    &                        &  median/mean   & median/mean\\
\hline
 [6:10:10:1]$_{\_181}$ &0.100           & 0.190/0.193        & 0.074/0.078  &[7:10:10:1]$_{\_201}$&0.090      &  0.163/0.171&0.065/0.065\cr
 [6:10:9:1]$_{\_179}$  &0.103           & 0.191/0.191        & 0.074/0.075  &[7:10:9:1]$_{\_189}$ &0.087       &  0.174/0.173&0.067/0.065\cr
 [6:10:8:1]$_{\_167}$  &0.103           & 0.193/0.203        & 0.074/0.079  &[7:10:8:1]$_{\_177}$ &0.083       &  0.166/0.172&0.066/0.067\cr
 [6:10:7:1]$_{\_155}$  &0.107           & 0.192/0.195        & 0.074/0.075  &[7:10:7:1]$_{\_165}$ &0.090      &  0.167/0.175&0.066/0.065\cr
 [6:10:6:1]$_{\_143}$  &0.107           & 0.191/0.203     & 0.076/0.079  &[7:10:6:1]$_{\_153}$ &0.090      &  0.162/0.174&0.063/0.066\cr
 [6:10:5:1]$_{\_131}$  &0.110           & 0.172/0.184     & 0.070/0.073  &[7:10:5:1]$_{\_141}$ &0.093      &  0.162/0.184&0.064/0.069\cr
 [6:9:5:1]$_{\_119}$   &0.110           & 0.183/0.200     & 0.074/0.078  &[7:9:5:1]$_{\_128}$  &0.097      &  0.155/0.171&0.058/0.061\cr
 [6:8:5:1]$_{\_107}$   &0.120           & 0.187/0.209      & 0.075/0.079  &[7:8:5:1]$_{\_115}$  &0.103      &  0.158/0.177&0.062/0.065\cr
 [6:7:5:1]$_{\_95}$    &0.120           & 0.190/0.214      & 0.075/0.080  &[7:7:5:1]$_{\_102}$  &0.103      &  0.147/0.161&0.056/0.058\cr
 [6:6:5:1]$_{\_83}$    &0.133           & 0.211/0.230      & 0.075/0.077  &[7:6:5:1]$_{\_89}$   &0.113      &  0.149/0.159&0.059/0.061\cr
 [6:5:5:1]$_{\_71}$    &0.153            & 0.216/0.227     & 0.076/0.078  &[7:5:5:1]$_{\_76}$   &0.130      &  0.140/0.155&0.057/0.060\cr
 [6:5:4:1]$_{\_64}$    &0.153           & 0.233/0.247     & 0.080/0.083  &[7:5:4:1]$_{\_69}$   &0.130      &  0.144/0.156&0.059/0.062\cr
 [6:5:3:1]$_{\_57}$    &0.167           & 0.263/0.290     & 0.086/0.093  &[7:5:3:1]$_{\_62}$   &0.137      &  0.159/0.170&0.062/0.064\cr
 [6:5:2:1]$_{\_50}$    &0.213           & 0.275/0.269      & 0.088/0.087  &[7:5:2:1]$_{\_55}$   &0.150      &  0.154/0.156&0.060.0.061\cr
 [6:5:1:1]$_{\_43}$    &0.303           & 0.291/0.290      & 0.126/0.125  &[7:5:1:1]$_{\_48}$   &0.240      &  0.194/0.195&0.074/0.074\cr
\hline
 [6:20:1]$_{\_161}$    &0.300           & 0.283/0.281     &0.117/0.118   &[7:20:1]$_{\_181}$   &0.237       &  0.226/0.225&0.086/0.084\cr
 [6:15:1]$_{\_122}$    &0.293           & 0.277/0.275      &0.116/0.118   &[7:15:1]$_{\_136}$   &0.230       &  0.228/0.229&0.087/0.085\cr
 [6:10:1]$_{\_81}$     &0.273           & 0.259/0.258      &0.105/0.106   &[7:10:1]$_{\_91}$    &0.223       &  0.207/0.219&0.083/0.084\cr
 [6:5:1]$_{\_41}$      &0.340           & 0.287/0.289     &0.117/0.120   &[7:5:1]$_{\_46}$     &0.273       &  0.261/0.259&0.097/0.097\cr
\hline
  $\dag$ [---] $_{\_83..215}$ &0.105        & 0.190/0.206     &0.071/0.075   & [---]$_{\_93..225}$ &0.086       &  0.175/0.186&0.065/0.065\cr
\hline
\hline
%% \multicolumn{9}{l}
\label{tab:architecture}
\end{tabular}
{$\dag$ Training and combination of different architectures (n:10:1..12:1). In
the second hidden layer the number of units ranges from 1 to 12. \hfill}
\end{table*}
\large

\subsection{Training on the HDF--N}
\subsubsection{Colors as input pattern}

 \begin{figure}
 \includegraphics[width=90mm]{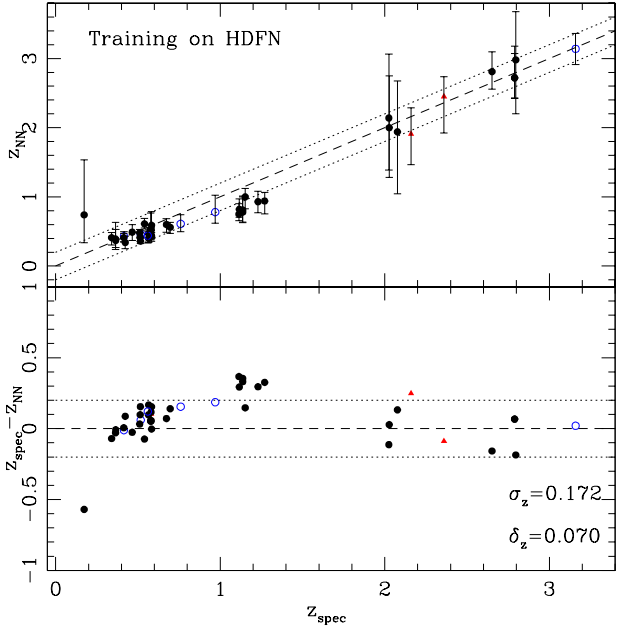}
 \caption{Comparison between spectroscopic redshift in the HDF--S and
 the neural redshift using the colors as an input pattern.  The training has been
 done on the HDF--N spectroscopic sample, the
 estimation of the redshift for each object is the median of 100
 predictions and the error bars represent 1-$\sigma$ interval.
 Open circles represent objects with unreliable photometry and triangles
 are objects with uncertain spectroscopic redshift.
\label{fig:BEST23}}
  \end{figure}

 \begin{figure}[h]
 \includegraphics[width=90mm]{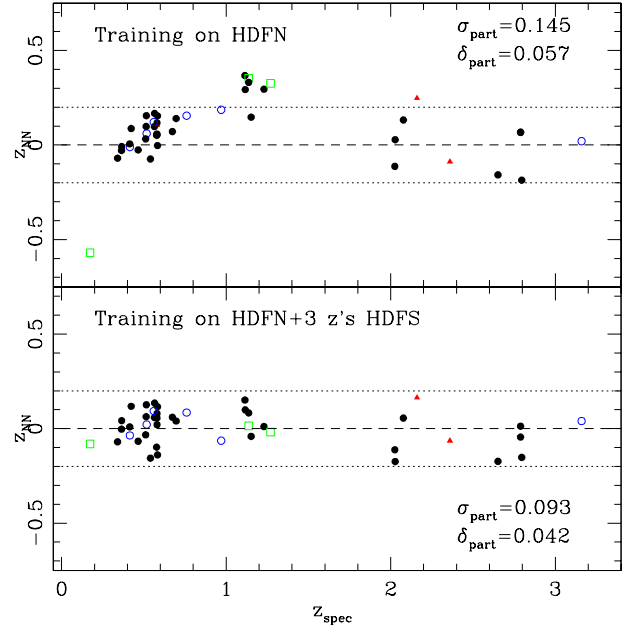}
 \caption{The effects of adding information. 
 Upper panel: comparison between spectroscopic redshift and the neural
 redshift for the spectroscopic sample in the HDF--S. The training has
been carried out on the HDF--N spectroscopic sample
(150 objects, as shown in the lower panel of the Fig.~\ref{fig:BEST23}).
The partial error ($\sigma_{part}$)
has been considered, i.e. the dispersion calculated without the three objects marked 
with the open square symbols, see text).  
 Lower panel: comparison between spectroscopic
 redshift and the neural redshift for the spectroscopic sample in the HDF--S,
the open squares symbols show the three objects that have been used during the training
(in addition to the 150 objects in the HDF--N), this new information improves the
partial error (i.e. the $\sigma_{part}$ calculated without these three objects), 
in particular at redshift around 1.2.
\label{fig:BEST23_SUGG}}
  \end{figure}

The input pattern contains the colors of the galaxies
($\frac{f_{F300}}{f_{F814}}$, $\frac{f_{F450}}{f_{F814}}$,
$\frac{f_{F606}}{f_{F814}}$, $\frac{f_{J}}{f_{F814}}$,
$\frac{f_{H}}{f_{F814}}$, $\frac{f_{K}}{f_{F814}}$), normalized as
described in Section 4.1.

The training has been carried out setting the maximum number of epochs
to 5000. The distribution of weights corresponding to the minimum
training error has been stored. We have verified that 5000
epochs are sufficient in this case to reach the convergence of the system.
Trainings done on 10000 and 15000 epochs give similar results. 

The dispersion $\sigma_{z}^{test}$ obtained for the spectroscopic sample in
the HDF--S is shown in Table~\ref{tab:architecture} (left
side). Different architectures have been used with one and two hidden
layers and different numbers of nodes.

The comparison between $zspec$ and $zNN$ for the architecture
6:10:5:1 (six input nodes, two hidden layers with ten and five
units and one output nodes) is shown in Fig.~\ref{fig:BEST23}. The
resulting error is $\sigma_{z}^{test}=0.172$. The systematic
errors are common to all the explored architectures. In particular
there is a clear discrepancy for the object at $z=0.173$ (ID=667
in the Tables of \cite{vanzella01}), due to the insufficient
information available in that redshift regime. A systematic
underestimation for the group of objects at redshift around 1.2 is
also evident. Combining different architectures with different
numbers of units in the second hidden layer (from 1 to 12), the
result does not change, the dispersion in the test set is
compatible with the dispersion obtained using a fixed
architecture.

For networks with a low complexity the error ($\sigma_{z}^{test}$)
starts increasing together with the $<\sigma_{train}>$ (the $<\sigma_{train}>$
is the mean of the training errors ($\sigma_{train}$) obtained in the bootstrap).
The same happens with networks with one hidden layer (see Table~\ref{tab:architecture}).

These results show that, although one hundred extractions (100 members) are enough to
diminish the random errors, new information in the training set is needed in order to
reduce the systematic errors.

This is clearly shown in Fig.~\ref{fig:BEST23_SUGG} where  we have
added to the training set three objects belonging to the HDF--S
spectroscopic sample: ID=667 with the discrepant redshift
mentioned above and two objects randomly chosen from the group
around redshift 1.2. In the upper panel of
Fig.~\ref{fig:BEST23_SUGG} the square symbols represent these
three objects used in the training together with the 150 in the
north, the dispersion in the HDF--S is calculated on the rest of
the sample (41 objects, $\sigma_{part}$). The training on the 150
objects gives as prediction $\sigma_{part}$=0.145. By computing
the training in the same conditions but with 153 objects rather
than 150, the prediction around redshift 1.2 clearly improves, and
a $\sigma_{part}$=0.093 is obtained. The predictions for the rest
of the objects do not change significantly. The improvement for
the square symbols is obvious (it is due to the learning
algorithm).
The network shows a remarkable ability to learn the new signal
present in the training set.

In the next section the colors together with the apparent luminosity
in the $F814$ band will be used as input pattern.

\subsubsection{Colors and apparent luminosity as an input pattern}

 \begin{figure}
 \includegraphics[width=90mm]{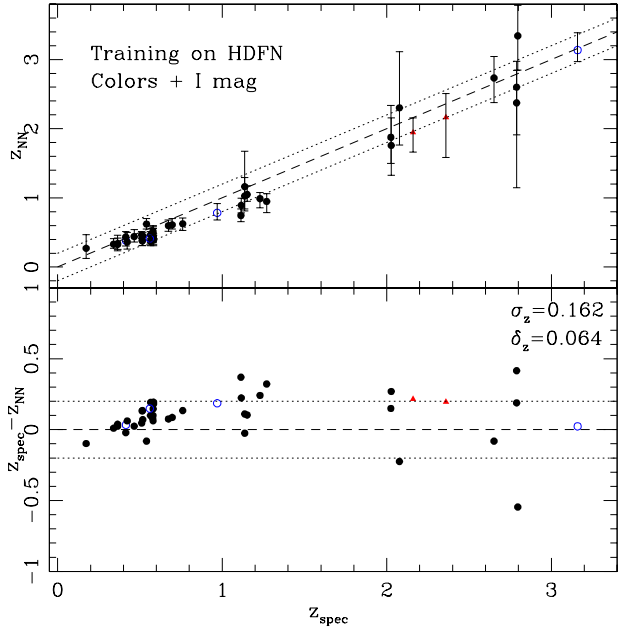}
 \caption{Comparison between spectroscopic redshift and the neural
redshift for the spectroscopic sample in the HDF--S. The training has
been carried out on the HDF--N spectroscopic sample, the estimation of
the redshift for each object is the median of 100 predictions.
The input pattern is composed of colors and the apparent luminosity in
the $F814$ band. The symbols are the same as in
 Fig.~\ref{fig:BEST23}.
\label{fig:BEST24}}
  \end{figure}
The input pattern contains the colors and the apparent
luminosity in the $F814$ band.
Also in this case we have performed one hundred training on the
150 galaxies in the HDF--N. The dispersion $\sigma_{z}^{test}$
obtained for the spectroscopic sample in the HDF--S is shown in
Table~\ref{tab:architecture} (right side, {\it ``colors $\&$
mag.''}).  In general, the predictions are better than the results
obtained with only colors as an input pattern. In this application
the magnitude information improves the prediction at low redshift
(in particular for the object ID=667).  On the other hand the
scatter at high redshift seems to increase, if compared with the
case with only colors as an input (see Fig.~\ref{fig:BEST24}).
There is still a bias (although reduced) at redshifts around 1.2.

The training on different architectures, 6:10:1:1, 6:10:2:1, ... and
6:10:12:1 (6:10:1..12:1 hereafter) produces a
dispersion similar to that obtained by fixing the architecture.  Also in
this case the networks with a low complexity %6:7:6:1 and 6:5:5:1
produce a large error both in the $<\sigma_{train}>$ and in the
$\sigma_{z}^{test}$. The same happens with networks with one hidden
layer (see Table~\ref{tab:architecture}).

These tests show that the information introduced by the apparent
luminosity produces a slight improvement: the error is always less
than the error obtained using only colors
(but the sample is still too small to generalize this result).

The problem concerning the completeness of the training set is common in the
empirical technique for the estimation of the redshift.
There is a well known gap without spectroscopic redshifts in the interval
(1.3,2) due to the absence of observational spectroscopic features.
Moreover, spectroscopic surveys are flux limited and
the spectroscopic redshifts tend to be available only for brighter objects.
To solve this problem and fill the above mentioned gap it is useful to
introduce in the training set examples derived from observed or
synthetic template SEDs.

\begin{table*}
%\onecolumn
\centering
\caption{Training of different architectures on the HDF--N
spectroscopic sample and a set of templates derived from CWWK.
The evaluation is on the HDF--S
spectroscopic
sample. The bootstrap has been computed on 100 extractions (100
members of the committee). In the ``training data'' column, ``+150''
means that the 150 spectroscopic redshifts in the HDF--N have been used in
addition to the CWWK SEDs.}
\begin{tabular}{lcccccccc}
\hline \hline
 [Net]$_{\_Weights}$   &  Epochs &Training&$dz$ &
E(B-V)&$<\sigma_{train}>$ &$\sigma_{z}^{test}$&$\delta_{z}^{test}$\\
                       &         &  Data  &     &       &                   &   median/mean     &
median/mean\\
\hline
 [6:30:30:1]$_{\_1171}$&1000     &3206+150&0.01 & 0 &0.059 &0.142/0.143 & 0.054/0.056 \cr
 [6:25:25:1]$_{\_851}$ &1000     &3206+150&0.01 & 0 &0.078 &0.132/0.133 & 0.057/0.056 \cr
 [6:20:20:1]$_{\_581}$ &1000     &3206+150&0.01 & 0 &0.062 &0.131/0.128 & 0.056/0.054 \cr
 [6:15:15:1]$_{\_361}$ &1000     &3206+150&0.01 & 0 &0.065 &0.135/0.128 & 0.058/0.055 \cr
 [6:15:10:1]$_{\_276}$ &1000     &3206+150&0.01 & 0 &0.064 &0.131/0.127 & 0.058/0.055 \cr
 [6:10:15:1]$_{\_251}$ &1000     &3206+150&0.01 & 0 &0.078 &0.128/0.127 & 0.060/0.059 \cr
 [6:10:10:1]$_{\_191}$ &1000     &3206+150&0.01 & 0 &0.076 &0.138/0.133 & 0.064/0.060 \cr
 [6:10:5:1]$_{\_131}$  &1000     &3206+150&0.01 & 0 &0.076 & 0.138/0.132 & 0.064/0.062\cr
 [6:7:6:1]$_{\_104}$   &1000     &3206+150&0.01 & 0 &0.106 & 0.159/0.157 & 0.076/0.075 \cr
 [6:5:5:1]$_{\_71}$    &1000     &3206+150&0.01 & 0 &0.173     &0.198/0.186 &0.084/0.080\cr
\hline \hline
 [6:20:20:1]$_{\_581}$ &500      &12824+150&0.01 & 0.0,0.05,0.1,0.2&0.060 &0.132/0.135&0.056/0.055 \cr
\hline
 [6:15:10:1]$_{\_276}$ &5000     &646+150 & 0.05& 0 & 0.068   & 0.125/0.127&0.057/0.057 \cr
 [6:15:10:1]$_{\_276}$ &5000     &326+150 & 0.1 & 0 & 0.093 & 0.127/0.128&0.059/0.060 \cr
\hline
 *[6:10:1..12:1]$_{\_83..251}$ &10000     &326+150 & 0.1 & 0 & 0.086 & 0.134/0.133& 0.062/0.062\cr
\hline
\multicolumn{8}{l}
{* Training on different architectures, in the second hidden layer the number of
units ranges from 1 to 12 (6:10:1..12:1).\hfill}
\label{tab:CWWl_BEST3}
\end{tabular}
\end{table*}

 \begin{figure}
 \includegraphics[width=90mm]{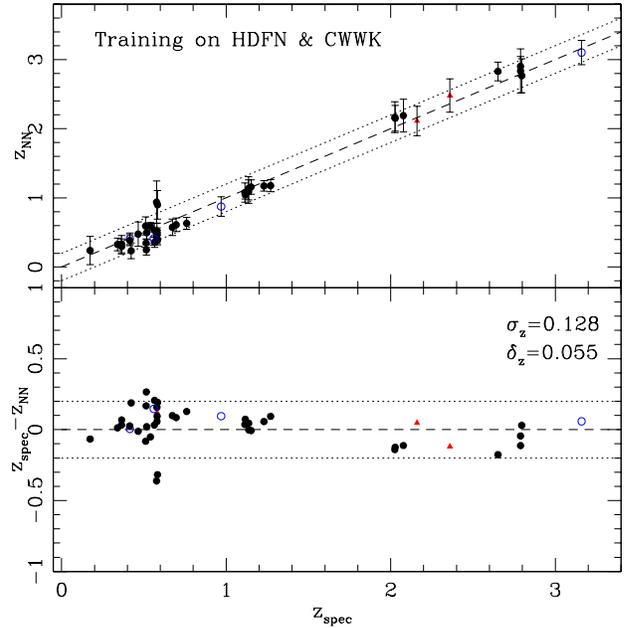}
 \caption{Comparison between spectroscopic redshift in the HDF--S and
 the neural redshift obtained with a committee of networks and using as
input pattern the colors.  The
 estimation of the redshift for each object is the mean of 100
 predictions and the error bars represent 1-$\sigma$ interval. The
 training set is composed by CWWK SEDs mixed with the spectroscopic
 sample in the HDF--N. The symbols are the same as in
 Fig.~\ref{fig:BEST23}. }
\label{fig:CWWl_6_15_15_1_3206_1000}
  \end{figure}

\subsection{Combination of training sets}

\subsubsection{Training on HDF--N mixed with CWW SEDs}
Increasing the information in the training data is an
obvious method to improve the generalization.

As a first approach to produce a complete range of galaxy SEDs we have
adopted the templates of Coleman, Wu \& Weedman (1980) for a typical
elliptical, Sbc, Scd and Irregular galaxy plus two
spectra of star-forming galaxies
(SB1 and SB2 from the atlas of \cite{kinney96}).
This choice is similar to the approach of \cite{soto99} and
\cite{arno99} and in the following will be referred to as ``CWWK''.

Galaxies have been simulated in the redshift range 0$<$z$<$6.
3206 SEDs have been drawn from the CWWK templates with a step in
redshift equal to 0.01 ($dz=0.01$). Extinction effects have been introduced
($E(B-V)=0.05, 0.1,0.2$) adopting a Calzetti extinction law (\cite{calzetti97}).
12824 SEDs have been produced in this way. The CWWK templates do not
take into account the evolution of galaxy SEDs with cosmic time.

A committee of 100 networks has been adopted and the median and mean
values have been used to estimate the redshift.

In Table~\ref{tab:CWWl_BEST3} the prediction for the HDF--S
spectroscopic sample is shown. A series of tests has been carried out
both neglecting the effects of intrinsic extinction and introducing an
extinction effect. No significant difference
in the predictions has been measured. The number of training data and
the $<\sigma_{train}>$ are also shown.

The predictions for the HDF--S are clearly improved taking into
account the information derived from the CWWK templates and remain
stable almost independently of the architecture
($\sigma_{z}^{test} \simeq 0.13$). 
Low complexity networks (6:7:6:1 and 6:5:5:1)
produce large errors: these are clear cases of under-fitting in the
training data. 
In Fig.~\ref{fig:CWWl_6_15_15_1_3206_1000} the comparison between the
spectroscopic redshifts and the neural predictions is shown for
the network 6:15:15:1
and bootstrap process. 
The prediction improves at redshift around 1 and for the object ID=667 at
$z=0.173$.
At high redshift ($z>2$) the uncertainty of the individual redshift estimates is
significantly reduced (compare, for example, the error bars at 
$z>2$ in Fig.~\ref{fig:BEST23} and in
Fig.~\ref{fig:CWWl_6_15_15_1_3206_1000}).  

Reducing the step in redshift ($dz$=0.01, 0.05, 0.1) and hence the
number of training data, leaves the prediction stable.  The
trainings computed on a reduced sample, 326+150 examples (326 CWWK
SEDs and 150 spectroscopic redshifts in the HDF--N) with $dz=0.1$
and 646+150 examples with $dz=0.05$ without extinction, give the
same result obtained with $dz=0.01$. This means that the committee
of networks is able to achieve the same fit in the color space also
with a reduced grid.
 \begin{figure}
 \includegraphics[width=90mm]{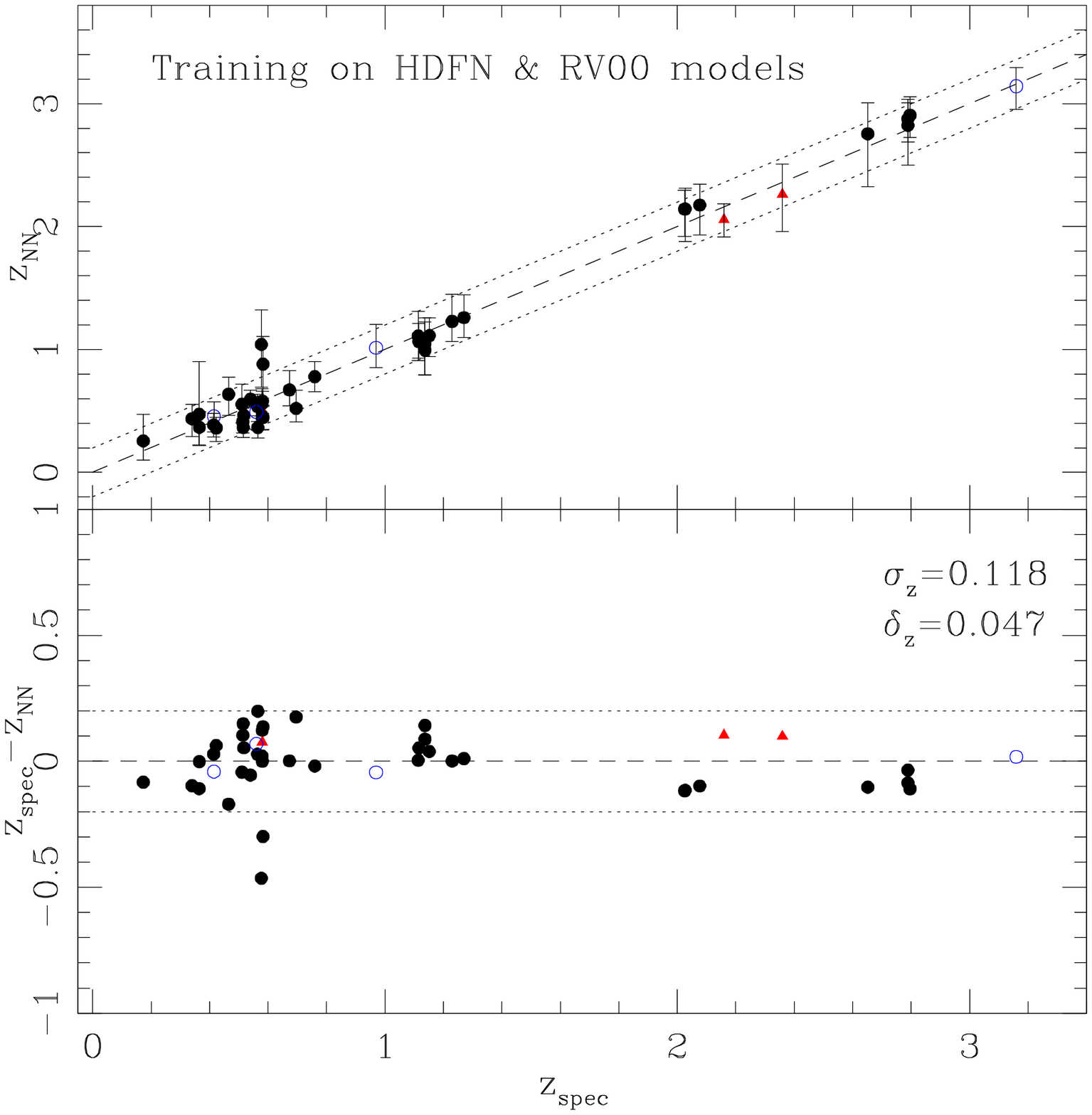}
 \caption{Comparison between spectroscopic redshift in the HDF--S and the
neural redshift obtained with a committee of networks and using as
input pattern the colors. The  estimation of the redshift for each
object is the mean of 100  predictions and the error bars
represent 1-$\sigma$ interval. The  training set is composed of
RV00 models  (bootstrap on 1000 RV00 SEDs and 150 spectroscopic
redshifts in the HDF--N, see Table~\ref{tab:RV00}). The symbols
are the same as in Fig.~\ref{fig:BEST23}.}
\label{fig:1000RV00e150HDFN_620201_2000}
  \end{figure}
\begin{figure*}
\includegraphics[width=170mm]{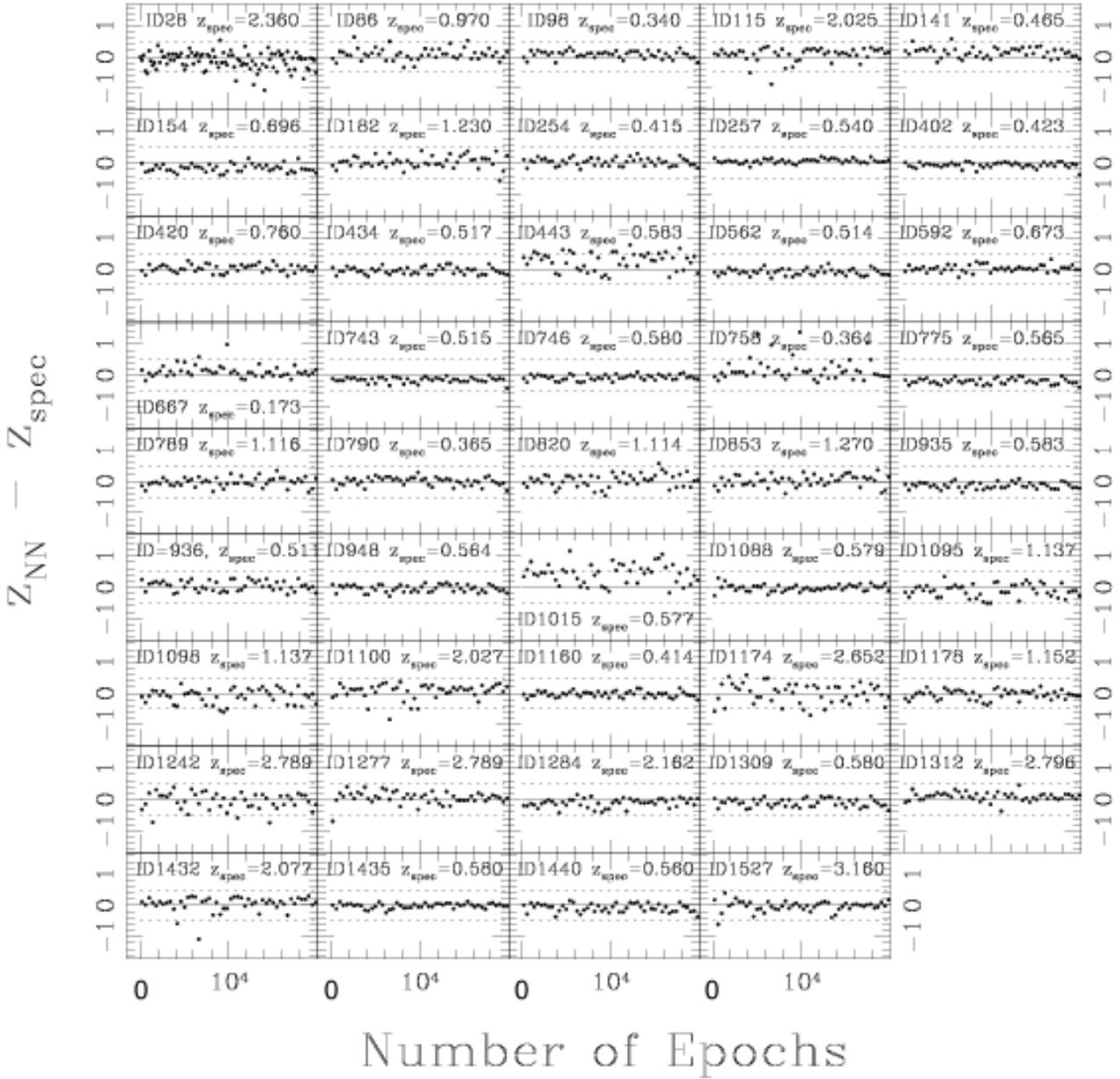}
 \caption{Predictions of a (6:20:20:1) NN for 44 galaxies in the HDF--S
 as a function of the epoch
 (an epoch correspond to the processing of all the examples one time, as
 defined in Sect.~2.1.2).
The training has been carried out
on the spectroscopic sample in the HDF--N and on RV00 templates, 
using as an input pattern the colors and the I mag. 
The ordinate shows the difference between the
prediction of the NN, $z_{NN}$, at a given epoch and the actual
spectroscopic redshift $z_{spec}$. The numbers in the upper left
part of the panels correspond to the galaxy identifiers in the
catalog by \cite{vanzella01}.} \label{fig:colors}
\end{figure*}
\subsubsection{Training on the HDF--N mixed with Pegase models}
We have also trained the neural system on the
HDF--N spectroscopic sample
and a set of models derived from the most recent version of the code by M. Fioc
and Rocca-Volmerange (\cite{rv97}),
named Pegase 2.0 (RV00).

In the Rocca--Volmerange code the star formation history is
parameterized by two $e-folding$ star formation time-scales, one
($\tau_g$) describing the time--scale for the gas infall on the
galaxy and the other ($\tau_*$) the efficiency of gas to star
conversion. By tuning the two time--scales it is possible to
reproduce a wide range of spectral templates, from early types (by
using small values of $\tau_g$ and $\tau_*$) to late types. For
the earliest spectral type, a stellar wind is also assumed to
block any star--formation activity at an age $t_{wind}$.  The
major advantage of the Rocca--Volmerange is that it allows to
follow explicitly the metallicity evolution, including also a
self--consistent treatment of dust extinction and nebular
emission.  Dust content is followed over the galaxy history as a
function of the on--going star--formation rate, and an appropriate
average over possible orientations is computed. Although more
model--dependent, this approach has the advantage of producing the
evolutionary tracks of several galaxy types with a
self--consistent treatment of the non-stellar components (dust and
nebular emission).
An application of the PEGASE 2.0 code to photometric redshifts
has been recently
presented by \cite{rv02}.

We have followed the training technique described in Sec. 3.1.2.
Adopting the scenarios described in \cite{rv02} we have obtained three 
samples from the RV00 package: 112824, 28544 and
14400 models with step in redshift $dz=0.025$, $dz=0.1$ and $dz=0.2$,
respectively (0$<z<$6). An other training sample has been obtained 
from the 112824 sample dimming the fluxes by a factor of 10 and 100
and considering as the training set the templates with apparent luminosity 
in the $F814$ band less than 27, in this way 201757 objects have been carried 
out.

\begin{table*}
%\onecolumn
\centering
\caption{Training of different architectures on the HDF--N
spectroscopic sample and a set of templates derived from Rocca Volmerange
(redshift in the interval $z=0-6$). The evaluation is on the HDF--S
spectroscopic
sample. The bootstrap has been computed on 100 extractions (100
members of the committee). In the training data column, ``+150''
means that the 150 spectroscopic redshifts in the HDF--N have been used in
addition to the RV00 models.}
\begin{tabular}{lccccccc}
\hline \hline
 [Net]$_{\_Weights}$   &  Epochs &Training&$dz$ &$<\sigma_{train}>$ &$\sigma_{z}^{test}$&$\delta_{z}^{test}$\\
                       &         &  Data  &     &                   &   median/mean     & median/mean\\
\hline
 [6:20:20:1]$_{\_581}$ &2000     &1000+150&0.025 &0.168 &0.118/0.120 & 0.047/0.050 \cr
 [6:20:20:1]$_{\_581}$ &3000     &300+150&0.1    &0.171 &0.123/0.119 & 0.054/0.053 \cr
 [6:20:20:1]$_{\_581}$ &5000     &150+150&0.2    &0.142 &0.123/0.116 & 0.053/0.051 \cr
\hline
 [6:30:30:1]$_{\_1171}$&2000     &1000+150&0.025 &0.167 &0.125/0.119 & 0.048/0.050 \cr
 [6:20:20:1]$_{\_581}$ &2000     &1000+150&0.025 &0.168 &0.118/0.120 & 0.047/0.050 \cr
 [6:10:10:1]$_{\_191}$ &2000     &1000+150&0.025 &0.176 &0.111/0.123 & 0.047/0.052 \cr
 [6:5:5:1]$_{\_71}$    &2000     &1000+150&0.025 &0.223 &0.159/0.164 & 0.064/0.068\cr
\hline
\hline
\label{tab:RV00}
\end{tabular}
\end{table*}

In the training on mixed samples the RV00 templates produce
slightly better results than the CWWK SEDs. A bootstrap process of
100 extractions has been carried out: at each extraction a random
sequence of the input patterns and a random initialization of the
weights have been adopted. At each extraction the training has
been computed on a set of data composed by 150 spectroscopic
redshifts in the HDF--N and a subset of models extracted randomly
from the RV00 samples. The performance in the south sample
is $\sigma_{z}^{test} \simeq 0.12$ (see Table~\ref{tab:RV00}).

Fig.~\ref{fig:colors} shows that no significant trend is 
present over the epochs
varying the initial distribution of weights and the sequence of the
training data (in the abscissa the epochs and in the ordinate the difference
$zNN-zspec$).
The prediction of the network becomes stable after
the first epochs (greater than 500) until the maximum epoch
(20000). The spread in the plots gives an indication of the resulting
uncertainty (also the spread is stable over the epochs).

Adopting a training set composed of RV00, CWWK
and the spectroscopic sample in the HDF--N produces a
$\sigma_{z}^{test} \simeq 0.12$,
of the same order of the dispersions obtained
with RV00+HDF--N and CWWK+HDF--N as training sets.

\subsubsection{Training on CWWK or RV00 templates}
\begin{table*}
%\onecolumn
\centering \caption{Training with various NN architectures on
templates derived from CWWK and RV00. The bootstrap has been
computed on 100 extractions (100 members of the committee).}
\begin{tabular}{lccccccccc}
\hline \hline
 [Net]$_{\_Weights}$   &  Epochs &Training&$dz$ & E(B-V)&$<\sigma_{train}>$ &$\sigma_{z}^{test}$&$\delta_{z}^{test}$&sample\\
                       &         &  Data  &     &       &                   &   median/mean     & median/mean      & \\
\hline
 [6:20:20:1]$_{\_581}$ &1000     &3206$_{\ CWWK}$    &0.01 & 0 & 0.036 & 0.180/0.186&0.068/0.067 & HDF--S \cr
 [6:20:20:1]$_{\_581}$ &500      &12824$_{\ CWWK}$   &0.01 & 0,0.05,0.1,0.2&0.044&0.196/0.200&0.067/0.068 & HDF--S\cr
\hline
 [7:20:20:1]$_{\_601}$ &10       &201757$_{\ RV00}$  &0.025& - & 0.157 & 0.153/0.158&0.068/0.070 & HDF--S\cr
 [7:20:20:1]$_{\_601}$ &10       &201757$_{\ RV00}$  &0.025& - & 0.157 & 0.259/0.257&0.061/0.062 & HDF--N\cr
 [7:20:20:1]$_{\_601}$ &10       &201757$_{\ RV00}$  &0.025& - & 0.157 & 0.231/0.233&0.064/0.064 & HDF--N/S\cr
\hline
%%\multicolumn{8}{l}
\label{tab:onlyTemplates}
\end{tabular}
\end{table*}
 \begin{figure}
 \includegraphics[width=90mm]{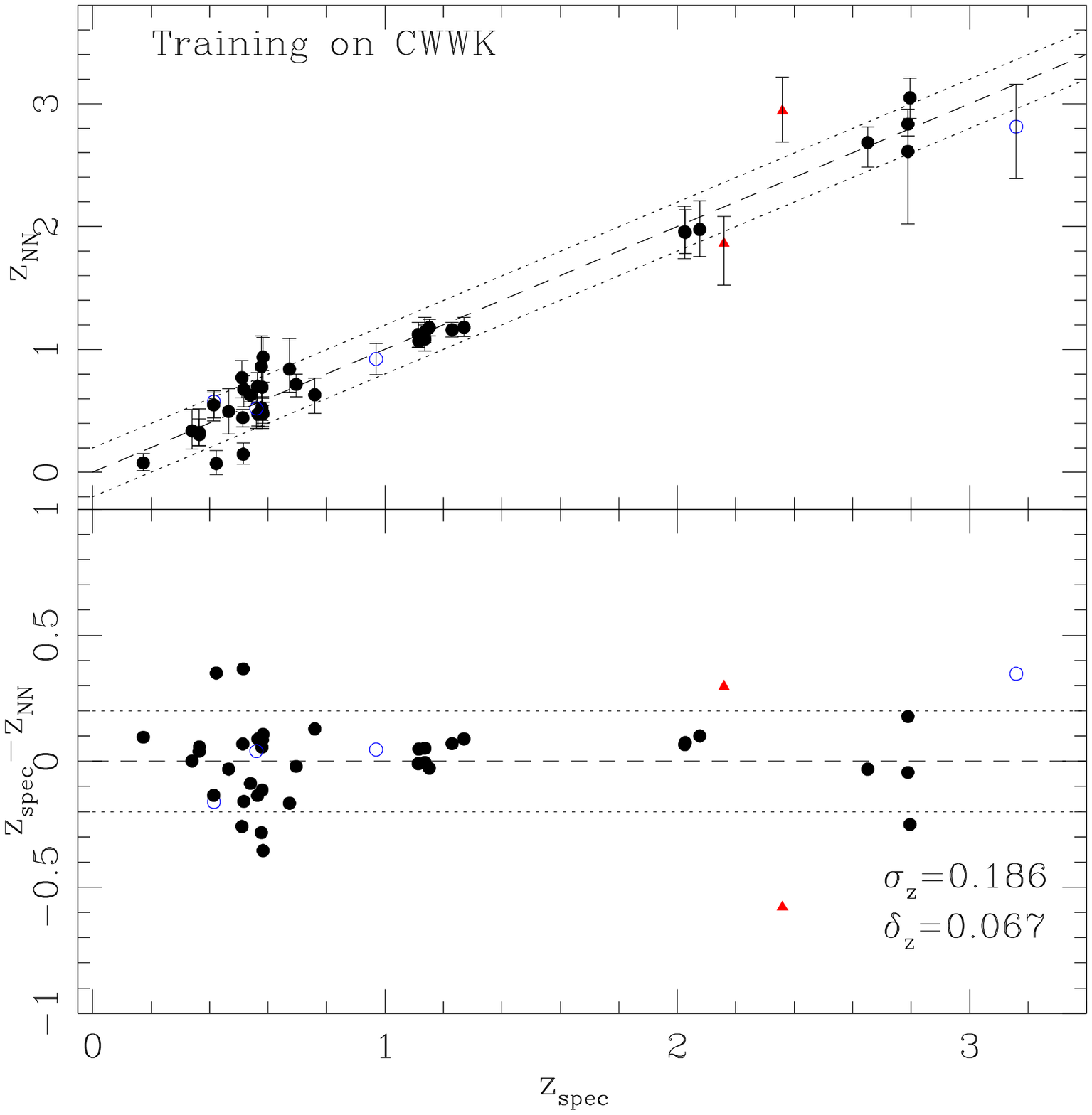}
 \caption{Comparison between spectroscopic redshift in the HDF--S and the
neural redshift obtained with a committee of networks and using as input
pattern the colors. The  estimation of the redshift for each object is the mean of
100  predictions and the error bars represent 1-$\sigma$ interval. The  training
set is composed of CWWK SEDs (3206 SEDs, see
Table~\ref{tab:CWWl_BEST3}). The symbols are the same as in
 Fig.~\ref{fig:BEST23}. }
\label{fig:OnlyCWWl_6_20_20_1_3206}
  \end{figure}

 \begin{figure}
 \includegraphics[width=90mm]{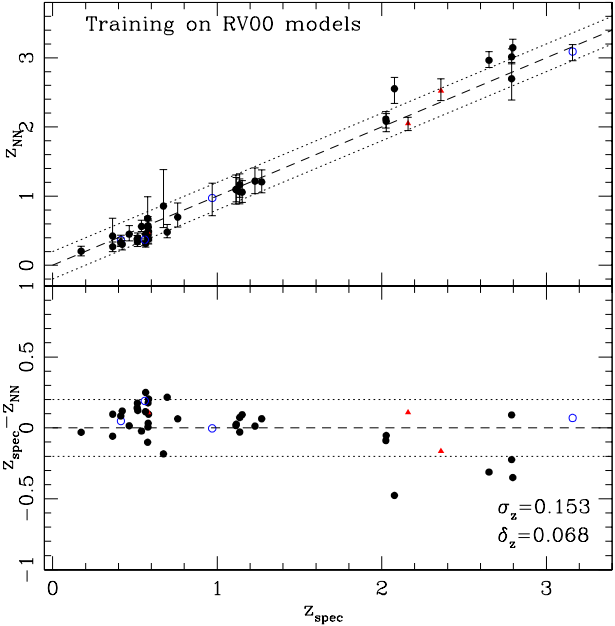} 
 \caption{Comparison between spectroscopic redshift in the HDF--S and the
neural redshift obtained with a committee of networks and using as input
pattern the colors. The  estimation of the redshift for each object is the mean of
100  predictions and the error bars represent 1-$\sigma$ interval. The  training
set is composed of RV00 models (112824 SEDs, see
Table~\ref{tab:RV00}). The symbols are the same as in
 Fig.~\ref{fig:BEST23}. }
\label{fig:onlyRV00_620201_112824_40}
  \end{figure}
Table \ref{tab:onlyTemplates} summarizes the results of various
trainings carried out only on templates, without the spectroscopic
redshifts.

Training on the colors derived from the CWWK templates produces a
dispersion in the HDF--S sample $\sigma_{z}^{test}=0.186/0.180$
(mean/median) (see Fig.~\ref{fig:OnlyCWWl_6_20_20_1_3206}). A
redshift step $dz=0.01$ and an extinction $E(B-V)=0.0$ were
adopted (3206 SEDs in the training set). A bootstrap on 100
extractions with maximum number of epochs set to 1000 was carried
out. Again, introducing the effects of extinction does not improve
this result.

Training on the colors and the apparent luminosity in the $F814$
band (7 inputs) derived from the RV00 models produces a dispersion
in the HDF--S sample $\sigma_{z}^{test}= 0.158/0.153$ (mean,
median), better than the estimates obtained with the CWWK SEDs.

\begin{figure}
 \includegraphics[width=90mm]{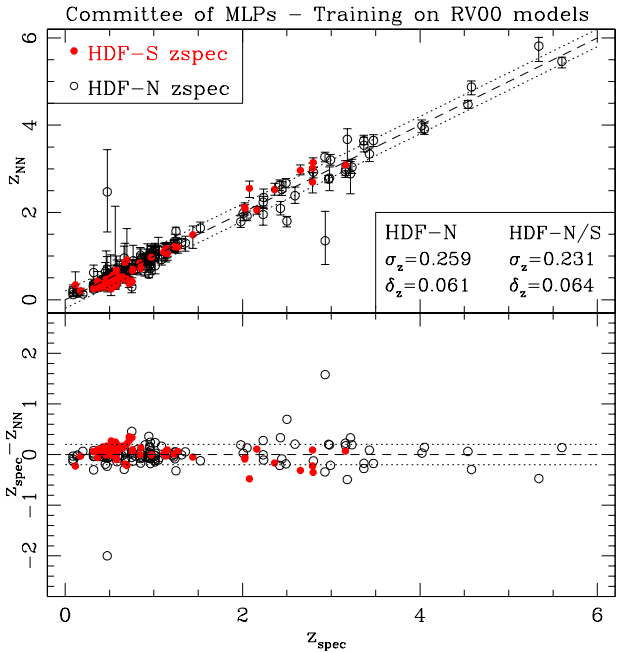} 
 \caption{Comparison between spectroscopic redshifts (HDF--N and HDF--S) and
the neural redshifts obtained with a committee of networks, using
as an input pattern the colors and the apparent luminosity in the
$F814$ band derived from RV00 models. The estimation of the
redshift for each object is the median of 100 predictions and the
error bars represent the 1-$\sigma$ interval.
} \label{fig:RV00HDF}
  \end{figure}

Fig.~\ref{fig:RV00HDF} compares the prediction of a NN trained on
the RV00 templates with the spectroscopic redshifts in the HDF--N
and HDF--S. The dispersion turns out to be $\sigma_{z}^{test} =
0.231$ for the full HDF--N plus HDF--S sample and $0.259$ for the
HDF--N only.

\begin{table}
%\onecolumn
\centering \caption{Summary of the different tests performed
on the HDF--S spectroscopic sample ($z<3.5$, 44 objects) described
in Section 5. The dispersion $\sigma_{z}$ is calculated in a low
redshift regime $z<2$ (34 objects) and high redshift regime $z>2$
(10 objects). }
\begin{tabular}{lcccc}
\hline \hline
 Training set    & $\sigma_{z}~(z<3.5)$ & $\sigma_{z}~(z<2)$& $\sigma_{z}~(z>2)$\\
                 &    44 objs.            &    34 objs.        &   10 objs.       \\
\hline
  HDF--N         &  0.172 & 0.186 & 0.114 \cr
  HDF--N mag.    &  0.162 & 0.139 & 0.222 \cr
  CWWK \& HDF--N &  0.128 & 0.131 & 0.114 \cr
  RV00 \& HDF--N &  0.118 & 0.128 & 0.094 \cr
  CWWK           &  0.186 & 0.146 & 0.282 \cr
  RV00           &  0.153 & 0.115 & 0.237 \cr
\hline
\label{tab:summary}
\end{tabular}
\end{table}
In Table~\ref{tab:summary} the tests on the HDF--S
spectroscopic sample are summarized. The dispersion is calculated
for 44 objects at $z<3.5$ and separately in the low-redshift
($z<2$) and high-redshift ($z>2$) regimes. In general the
performance improves when the information in the training set
increase.

\section{Application to the SDSS DR1}
\begin{figure}
 \includegraphics[width=90mm]{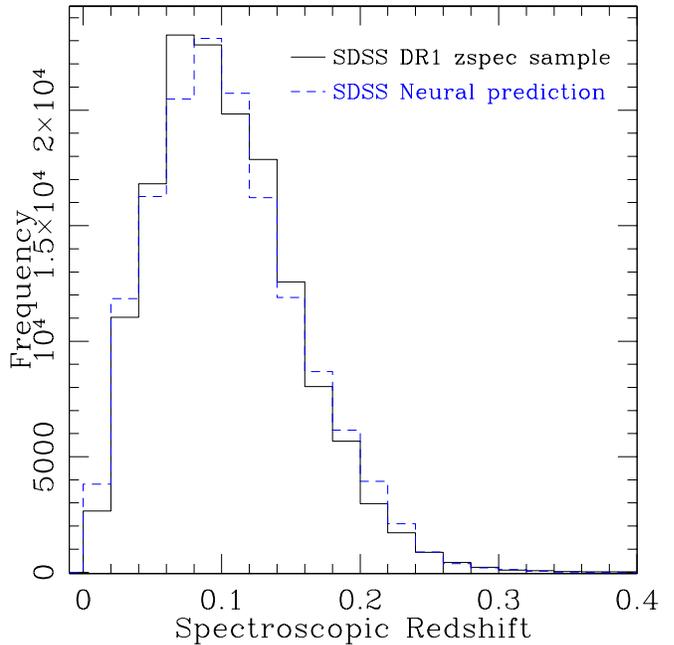}
 \caption{Redshift distribution of the spectroscopic sample obtained from the
 SDSS DR1 (113000 galaxies, solid line). The dashed line represents the distribution
 of the neural redshift prediction of the test sample (88108 galaxies) normalized to the
 total sample obtained with a 19:12:10:1 architecture (see text).
\label{fig:z_distr_SLOAN}}
\end{figure}
The Sloan Digital Sky Survey
\footnote[2]{Funding for the creation and
distribution of the SDSS Archive has been provided by the Alfred
P. Sloan Foundation, the Participating Institutions, the National
Aeronautics and Space Administration, the National Science Foundation,
the U.S. Department of Energy, the Japanese Monbukagakusho, and the
Max Planck Society. The SDSS Web site is {\tt
http://www.sdss.org/}. The Participating Institutions are The
University of Chicago, Fermilab, the Institute for Advanced Study, the
Japan Participation Group, The Johns Hopkins University, the
Max-Planck-Institute for Astronomy (MPIA), the Max-Planck-Institute
for Astrophysics (MPA), New Mexico State University, Princeton
University, the United States Naval Observatory, and the University of
Washington.}
(SDSS; \cite{york00}) consortium has publicly
released 134015 spectroscopic redshifts (\cite{abaza03}). The photometry in
the $ugriz$ bands and various image morphological parameters are also
available.

Recently, \cite{tagliaferri02} and \cite{firth02}
have used neural networks to produce photometric redshifts based on
the SDSS Early Data Release (SDSS EDR, \cite{stoug02}), while
\cite{ball03} have applied neural networks to the DR1 sample.

We have selected the data with the following criteria (see also
\cite{firth02}): (1) the spectroscopic redshift confidence must be
greater than 0.95 and there must be no warning flags, (2)
$r<17.5$. Moreover we have adopted the photometric criteria
proposed in Yasuda at al. (2001) for the star-galaxy separation.
An object is classified as a star in any band if the model
magnitude and the PSF magnitude differ by no more than 0.145. The
resulting catalog is almost entirely limited to $z<0.4$. The
redshift distribution of the DR1 sample is shown in
Fig.~\ref{fig:z_distr_SLOAN}.

\begin{figure}
 \includegraphics[width=90mm]{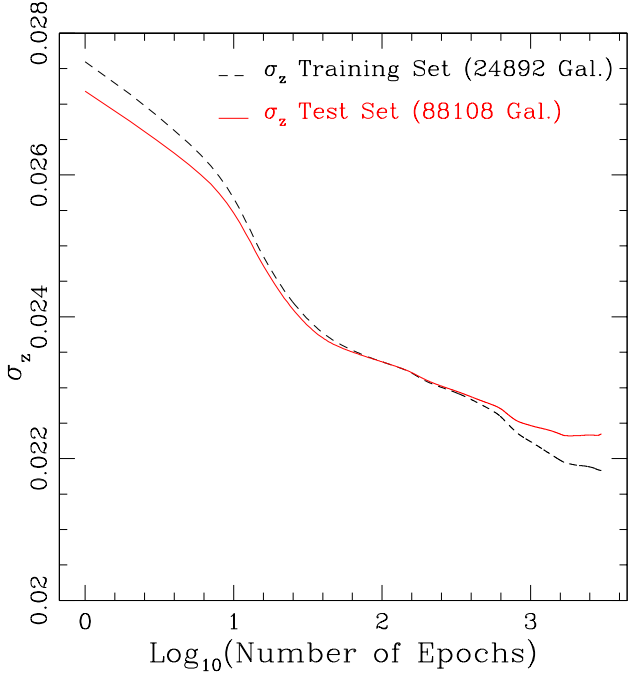}
 \caption{Behavior of the prediction as a function of the epochs for the
SDSS DR1 sample. The non-uniform
training sample has been used with the 19:12:10:1 architecture.
3000 epochs have been computed, the training and test errors are
shown as a function of the epoch.}
\label{fig:zTestError_71071_UNIF}
\end{figure}

Two different approaches have been explored in the NN
estimation of the DR1 photometric redshifts:
\begin{enumerate}
\item{
A 7:12:10:1 network with 3000 epochs and
10 different trainings, carried out
changing the initial random distributions of
weights and the sequence of the training examples.
The ``best'' distribution of weights corresponds to the lowest error
in the training sample (in almost all cases coincident with the last
epoch). The 7 input nodes are: the colors, the $r$-band magnitude,
the Petrosian 50 and 90 per cent $r$-band flux radii
($u-g$, $g-r$, $r-i$, $i-z$, $r$, $PetR50$, $PetR90$).
}
\item{
A 19:12:10:1 network with 15000 epochs and a single training carried
out. The additional inputs are in this case
the u-, g-, i-, z-band magnitudes
and the Petrosian 50 and 90 per cent flux radii in these bands.
}
\end{enumerate}
The results in terms of dispersions ($\sigma_{z}$ and
$|\Delta_{z}|$) and mean offsets $<\Delta_{z}>$ are summarized in
Table~\ref{tab:TAB_SLOAN.DR1}.
\begin{table*}
%\onecolumn
\centering
\caption{
{\it SDSS - DR1}: Training on 24892 galaxies (uniform and random sample). Test
on 88108 galaxies. The mean values are derived from 10
trainings
by varying the initial random distribution of weights and the sequence of
the training examples. In the first 2 rows 7 inputs nodes have been used (u-g, g-r, r-i, i-z, r, PetR50, PetR90).
Rows 3 and 4 correspond to a single training and 19 inputs have been used (u-g, g-r, r-i, i-z, u, g, r, i, z, PetU50,
PetU90, PetG50, PetG90, PetR50, PetR90, PetI50, PetI90, PetZ50, PetZ90).}
\begin{tabular}{lccccccccc}
\hline \hline
Net&  $\bf{W}$ & Epochs &Training&$<\sigma_{z}>$ & $<|\Delta_{z}|>$
&$<\Delta_{z}>$  &$<\sigma_{z}>$ & $<|\Delta_{z}|>$ & $<\Delta_{z}>$\\

    &           &        &  Data  &   (Train)     &      (Train)   &   (Train)  &   (Test)  &
(Test)   &    (Test)\\

\hline
 7:12:10:1 &273&3000&24892unif &0.026$\pm$0.0002 & 0.018$\pm$0.0002 &0.000 &0.024$\pm$0.0007&0.017$\pm$0.0005&0.004\\
 7:12:10:1 &273&3000&24892rand &0.023$\pm$0.0005 & 0.017$\pm$0.0004 &0.000 &0.023$\pm$0.0004&0.017$\pm$0.0004&0.002\\
\hline
Net&  $\bf{W}$ & Epochs &Training&$\sigma_{z}$ & $|\Delta_{z}|$
&$<\Delta_{z}>$  &$\sigma_{z}$ & $|\Delta_{z}|$ & $<\Delta_{z}>$\\
\hline
 19:12:10:1&381&15000&24892unif &0.025            & 0.017            &0.002 &0.023           &0.017           &0.001\\
 19:12:10:1&381&15000&24892rand &0.021            & 0.016            &-0.001&0.022           &0.016           &-0.002\\
\hline
\hline
\label{tab:TAB_SLOAN.DR1}
\end{tabular}
\end{table*}
Increasing the number of input nodes and the number of epochs improves
only slightly the result. In particular, Fig.~\ref{fig:zTestError_71071_UNIF}
shows the behavior of the training error for the 19:12:10:1 network
as a function of the ``current'' epoch,
shown until the maximum epoch 3000.
It is worth noting that, because of the {\it incremental learning}
method used in the present work (see Sect. 2.1.2), each epoch
corresponds to a number of variations of weights equal to the number
of training examples in the training set.  This explains why the
predictions of the network are good also at the very beginning (epoch
1) of the training phase.

\begin{figure}
 \includegraphics[width=90mm]{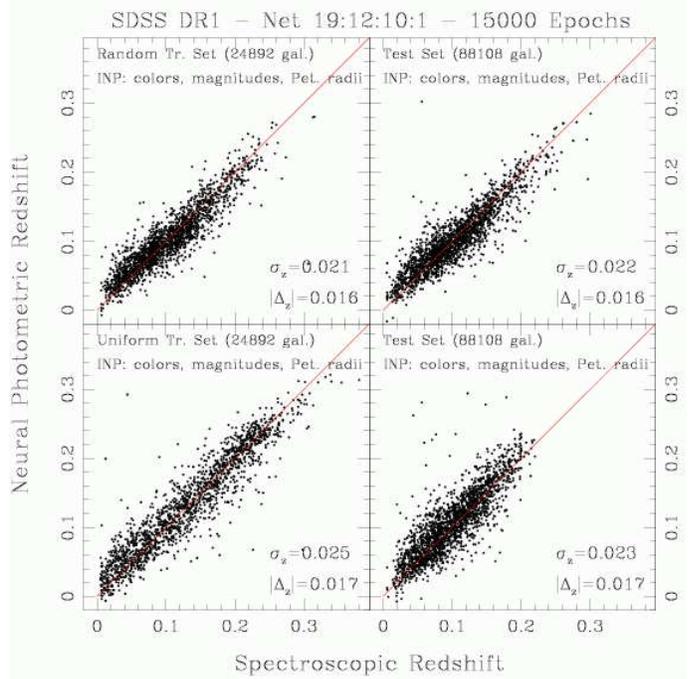}
 \caption{Redshift prediction in the SDSS DR1 (113000 galaxies) spectroscopic sample using a
19:12:10:1 architecture, 3000 epochs and 19 inputs (u-g, g-r, r-i, i-z, u, g, r, i, z, PetU50,
PetU90, PetG50, PetG90, PetR50, PetR90, PetI50, PetI90, PetZ50, PetZ90) as input pattern.
In the lower panel (training set on the left, test set on the right)
the training set has been built adopting a grid with a fixed step $dz$=0.000012
and extracting one galaxy for each interval of the grid (24892 galaxies
in total). In the upper panel (training set on the left, test set on the right)
the training set has been built extracting randomly a sample of
the same size (24892 galaxies) of the uniform sample.
In left panels only one point every 16 is plotted, while in the right panels only
a point every 50 is plotted.}
\label{fig:19INP_SLOAN.DR1}
\end{figure}

The highly inhomogeneous distribution of the redshifts (see
Fig.~\ref{fig:z_distr_SLOAN}) is expected to produce a bias in the
estimates, as discussed in \cite{tagliaferri02}, since any network
will tend to perform better in the range where the density of the
training points is higher. To investigate this effect two types of
training have been carried out: on a uniform training set and a
randomly extracted training set. The random and the uniform
training sets are both made of 24892 galaxies. In the cases of
randomly extracted training sets (Fig.~\ref{fig:19INP_SLOAN.DR1}
upper panels), a trend in the training and test phase is evident.
It appears as a distortion around z$\simeq$0.1, corresponding to
the higher density of training points (see
Fig.~\ref{fig:z_distr_SLOAN}). The behavior of the diagram using a
uniform training set is more regular
(Fig.~\ref{fig:19INP_SLOAN.DR1} lower panels).

Due to the large amount of data available, the trainings with and
without the validation set have produced indistinguishable results.
Also the dispersion obtained with a committee of networks and with a single
member is comparable, therefore no regularization has been applied and
a single training has been adopted in all cases.

Increasing the number of connections in the architecture
of the network does not cause the results to change significantly.  It
is interesting to note that even with a simple network 7:2:5:1 (34
weights and 7 input neurons), the dispersion obtained is comparable to
the 381 weights net (19:12:10:1). The 7:2:5:1 gives
$\sigma_{z}\simeq0.027$ ($|\Delta_{z}|\simeq0.021$) in the 88108 test
galaxies sample.

Various photometric redshift techniques (template-fitting,
Bayesian method, polynomial fitting, nearest-neighbor etc.) have
been applied to a similar spectroscopic sample extracted from the
SDSS EDR (see \cite{csabai02}). They produce in general
significantly worse results in terms of redshift dispersion,
except for the ``Kd-tree'', which shows a $\sigma_{z}=0.025$.

\begin{figure}
 \includegraphics[width=90mm]{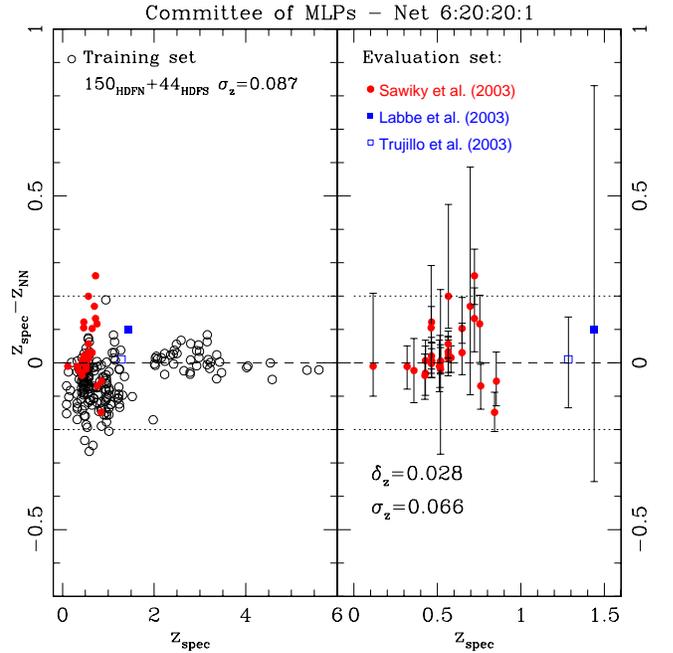}
 \caption{Comparison between spectroscopic redshift in the HDF--S and the
neural redshift obtained with a committee of networks and using as
input pattern the colors. The  estimation of the redshift for each
object is the mean of 100  predictions and the error bars
represent 1-$\sigma$ interval. In the left panels, the  training
set is composed of 150 (HDF--N) and 44 (HDF--S) spectroscopic
redshifts (open circles). The evaluation has been done on the
recent sample of spectroscopic redshifts (z$<$1) provided by
\cite{sawicki03}, filled circles, and on the large spiral galaxy
at z=1.439, square filled symbol (\cite{labbe03}) and on the
galaxy at z=1.248, open square symbol (\cite{tru03}). In the right
panels only the evaluation symbols are shown.}
\label{fig:150HDFN.44HDFS.10000.620201}
\end{figure}

\section{Summary and conclusions}
We have presented a new technique for the estimation of redshifts based
on feed-forward neural networks. The neural architecture has been
tested  on a spectroscopic sample in the HDF--S
(44 objects) in the range $0.1<z<3.5$
and on a large sample ($113000$ galaxies) derived from the SDSS DR1.

The flexibility offered by NNs
allows us to train the networks on sets that are
homogeneous (i.e. on spectroscopic redshifts or simulated templates)
or mixed (e.g. on spectroscopic redshifts and simulated data).
The galaxy templates for the training of the NNs with simulated data
have been derived from observational samples
(the CWWK SEDs) and from theoretical
data (P\'egase models).

The training on the theoretical data (colors and I mag. as 
input pattern) produces a $\sigma_{z}^{test}$ in the HDF--S
of the order of 0.15 (RV00),
while the training on the HDF--N
spectroscopic sample produces $\sigma_{z}^{test} \simeq 0.18$
(colors as input pattern) and $\sigma_{z}^{test} \simeq 0.15$
(colors and apparent $I$ luminosity as input pattern).
The training on mixed samples (observed SEDs with spectroscopic
redshift (HDF--N) and theoretical SEDs (CWWK or RV00 models))
improves the prediction, and a dispersion of the order of
$\sigma_{z}^{test} \simeq 0.11$ is reached.

At the end of the training
the NN contains ``experience'' that is a combination of the observed data
and the models.

It is interesting to note that the spectroscopic sample in the HDF--S can
be used either as a part of the training set
or as a validation set in order to
calibrate and tune the prediction (at least for the brighter objects)
and that with the increasing availability of spectroscopic redshift
the prediction can be continually improved.
As an example we have used both the HDF--N and the HDF--S
spectroscopic samples (194 objects in total) to predict with a 6:20:20:1
architecture the redshifts of $33$ galaxies in the range $0.1<z<1.5$
recently published by \cite{sawicki03}, \cite{labbe03} and
\cite{tru03}.
The resulting dispersion turns out to be $\sigma_{z} = 0.066$
(Fig.~\ref{fig:150HDFN.44HDFS.10000.620201}).

A reference dataset estimating photometric redshifts in the HDF--S
down to $I_{AB}\simeq 27$ has been produced: the training has been
performed on a set composed of RV00 models, 150
spectroscopic redshifts in the HDF--N and 77 spectroscopic
redshifts in HDF--S.

The better generalization obtained using a committee of networks with
respect to a single network is more evident in the case of small
training sets (Sec. 5.1 and 5.2). If the training
set is sufficiently complete and representative, good generalization
can be achieved also with a single training.

In summary the NN approach introduces the following advantages:

\begin{enumerate}
\item{Rapidity in the evaluation phase with respect to more
conventional techniques and possibility to deal with very large datasets.
The redshifts of $10^5$ galaxies can be estimated in
few seconds (using a laptop with PIII, 1.1 GHz).}
\item{The system can quickly learn new information, for example when
new spectroscopic
redshifts become available.}
\item{A priori knowledge (such as morphological properties, apparent
luminosity, etc.) can be taken into account.}
\item{There are no assumptions concerning the distribution of the input
variables.}
\item{Feed-forward NNs can also be implemented via hardware, in the so called
{\it machine learning} scheme. Neural processors have the same generalization
and learning ability as the MLP simulated via software (\cite{battiti95}), but
with an extremely high velocity performance ($10^{6-7}$ galaxies per second,
a very useful feature in the training phase).}
\end{enumerate}

Future developments include a better treatment of photometric
errors and upper limits, and the recognition of characteristics of the
galaxies (e.g. the type) from the input colors and/or morphological
features (such as the $Sersic$ index, luminosity profiles, etc.).

\begin{acknowledgements}
We warmly thank Walter Vanzella and Felice Pellegrino for the helpful
discussions about the NNs and the training techniques.
We thank Massimo Meneghetti for the useful discussions about the
training on the synthetic catalogs and Pamela Bristow for her
careful reading of the manuscript. This work was partially
supported by the ASI grants under the contract number ARS-98-226
and ARS-96-176, by the research contract of the University of
Padova ``The High redshift Universe: from HST and VLT to NGST''
and by the Research Training Network ``The Physics of the
Intergalactic Medium'' set up by the European Community under the
contract HPRN-C12000-00126 RG29185. EV thanks Lare for her
patience.

\end{acknowledgements}

\clearpage

\end{document}